\documentclass[preprint,aps]{revtex4}
\usepackage{graphicx}
\usepackage{dcolumn}
\usepackage{bm}
\usepackage{amsmath}
\begin{document}

\preprint{}
\title{Interacting two helical edge modes in quantum spin Hall systems}
\author{Yukio Tanaka$^{1}$, and Naoto Nagaosa$^{2,3}$}
\affiliation{$^1$Department of Applied Physics, 
Nagoya University, Nagoya, 464-8603,
Japan \\ 
$^2$ Department of Applied Physics, University of Tokyo, Tokyo 113-8656, Japan \\
$^3$ Cross Correlated Materials Research Group (CMRG), ASI, RIKEN, WAKO 351-0198, Japan 
}
\date{\today}

\begin{abstract}
We study theoretically the two interacting one-dimensional helical modes at the edges of the 
quantum spin Hall systems. 
A new type of inter-edge correlated liquid (IECL) without the spin gap is found. This liquid shows the diverging density wave (DW) and superconductivity (SC) 
correlations much stronger than those of the spinfull electrons. 
Possible experimental observations are also discussed.
\end{abstract}

\pacs{71.10.Pm,72.15.Nj,85.75.-d}
\maketitle



%

%




Quantum spin Hall system (QSHS) is a new state of matter 
topologically distinct from the 
usual band insulator~\cite{Mele,Bernevig,Bernevig2,Konig,Onoda}. 
It is protected by the band gap induced by the relativistic
spin-orbit interaction, and a $Z_2$ topological number characterizes it~\cite{Mele,Fu}. 
The simplest picture for QSHS is the two copies of quantum Hall system 
of up and down pseudospins with the opposite chiralities, and hence the edge modes 
with the opposite direction of the 
propagation for different pseudospins along the boundary of the sample are
expected \cite{Wen}. Time-reversal symmetry and Kramer's theorem guarantee the stability 
of these helical edge modes, i.e., prohibiting the backward scattering between 
the time-reversal pair states and protecting the crossing of the energy dispersions~\cite{wu2006,xu2006}. 
These modes, called the helical edge modes, are experimentally observed 
through the quantized charge conductance $2e^2/h$ in the quantum well of HgTe/(Cd,Hg)Te~\cite{Konig}. 
These helical edge modes on one edge are the "half" of the spinfull 
one-dimensional electrons, because only one chirality is allowed for each 
pseudospin. 
Therefore, several nontrivial features are expected for this one-dimensional 
system such as the half-$e$ charge near the ferromagnetic domain wall~\cite{Qi} and the 
robustness against the umklapp scattering~\cite{wu2006,xu2006}. 
Several other theoretical aspects have been studied very recently 
~\cite{Maciejko,Hou,Strom}. \par

In this paper, we study theoretically a new 
state created by the fusion/recombination of two of the helical edge modes,
i.e., the two half spinfull electrons, which is different from the original 
one-dimensional electrons. This fusion can be realized by the 
interaction between the edges of the 
opposite sides of the sample, or one can even design the interacting helical 
edge modes with the opposite  helicity (Fig. 1(a)) or 
same helicity (Fig. 1(b)). 
For a moment, we consider the anti-parallel case (Fig. 1(a)), and
the translation to the parallel case (Fig. 1(b)) is straightforward
and will be discussed later.
%
We employ the bosonization method. 
There are four fields involved in this problem, i.e., 
right moving up-pseudospin electron density $\rho_{1 R \uparrow}$, 
and left moving down-pseudospin electron density 
$\rho_{1 L \downarrow}$, for the edge 1, and 
$\rho_{2 R \downarrow}$,
$\rho_{2 L \uparrow}$, for the edge 2.
These densities are related to the phase fields of electrons 
( which are defined below Eq.(4) ) $\phi_{1 \uparrow}$,
$\bar{\phi}_{1 \downarrow}$, $\phi_{2 \downarrow}$,
$\bar{\phi}_{2 \uparrow}$, respectively, as 
$\rho_{1 R \uparrow} = (1/\sqrt{\pi}) \partial_{x} \phi_{1 \uparrow}$ 
etc.~\cite{Solyom,Senechal,Giamarchi} (see Fig. 1).
This means the spatial variation of the phase corresponds to the 
accumulation/depletion of the electrons. 
\begin{figure}[htb]
\begin{center}
\scalebox{0.8}{
\includegraphics[width=6cm,clip]{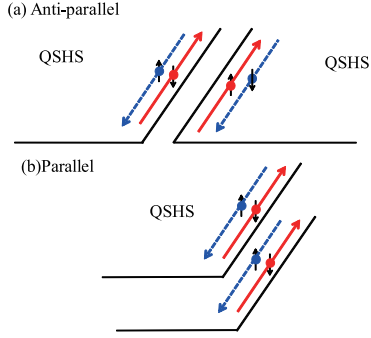}
}
\end{center}
\caption{
Schematic illustrations of two helical edge modes 
with (a)anti-parallel case and (b) parallel case. }
\label{fig:1}
\end{figure}
The novel states induced by the interaction is realized 
when some combination of these phases becomes rigid,
leading to the collective motion of the electrons.
We have found that the different combination of 
$\phi$'s gets rigid in the two interacting helical 
edge modes from those realized in the 
usual spinfull electrons. Namely, the helical edge modes 
offer an unique opportunity to realize a new state 
by the recombination of the "fraction" of electrons.   
To reach this conclusion, we have classified all the possible 
forms of the interaction, and found that the inter-edge pseudospin flip backward scattering 
plays an important role in the incommensurate case, while also the 
inter-edge umklapp scattering can be relevant in the commensurate case.
It is found that an inter-edge correlated liquid (IECL) analogous
to bipolaron liquid is realized in the incommensurate case, showing the density 
wave and superconducting correlations both of which 
are more strongly diverging compared with the one-dimensional spinfull 
interacting electrons. 
Furthermore, there is no spin gap in this liquid state, in sharp contrast to 
the Luther-Emery liquid or the spin gap state in ladder system. 
Physical properties of this novel liquid are described proposing 
possible experiments. \par

The electron-electron interaction is originally given by 
\begin{equation}
H_{\rm int} = { 1 \over 2} \int d r \int d r' 
\psi^\dagger_\sigma (r) \psi^\dagger_{\sigma'} (r')
V(r - r')
\psi_{\sigma'} (r') \psi_{\sigma} (r)
\label{eq:int}
\end{equation}
where $r,r'$ are the three-dimensional coordinates, 
$V(r-r')$ is the interaction potential, and 
the field operators $\psi^\dagger_{\sigma}(r)$, $\psi_\sigma(r)$ 
describe the creation and annihilation of the electron 
with spin $\sigma$ at the position $r$, respectively. 
Now we restrict to the helical edge modes only, which means
\begin{equation}
\psi_\sigma (r) \cong 
\sum_{\alpha=\uparrow, \downarrow} \sum_{a=1,2} 
c_{\alpha \sigma} \Psi_{a \alpha}(x) 
\varphi_a(y,z).
\end{equation}
Here, the index $a$ specifies the edge, $\sigma$ is the spin
of the electron, while $\alpha$ is the pseudospin in the 
presence of the spin-orbit coupling, and $c_{\alpha \sigma}$ 
is the coefficient of the transformation.
$\Psi_{a \alpha}(x)$ represents the field operator 
of the one-dimensional electrons along the edge channel, while 
$\varphi_a(y,z)$ is the wave function of the helical edge 
modes perpendicular to the edge $a$. 
We put this expression into Eq.(\ref{eq:int}), and 
neglect the overlap of the wave functions for the 
helical edge modes at different edges, which 
means that the different edges can not 
appear for the same
 spatial point $r$ or $r'$.
Also the pseudspin $\alpha$ determines the 
direction of the propagation for each edge.

When we consider the case where the direct overlap of the wave 
functions between the two edges can be neglected, 
the form of the interaction is restricted. 
The Hamiltonian $H= H_K + H_{\rm int}$ is the sum of the kinetic 
energy $H_K$ and the electron-electron interaction $H_{\rm int}$ as 
given by 
\begin{eqnarray}
H_K &=& -i\int dx [\Psi_{1\uparrow}^{\dagger}\partial_{x}\Psi_{1\uparrow}
- \bar{\Psi}_{1\downarrow}^{\dagger}\partial_{x}\bar{\Psi}_{1\downarrow}
\nonumber \\
&+& \Psi^{\dagger}_{2\downarrow}\partial_{x}\Psi_{2\downarrow}
- \bar{\Psi}_{2\uparrow}^{\dagger}\partial_{x}\bar{\Psi}_{2\uparrow}]
\label{eq:Hk}
\end{eqnarray}
and 
\begin{eqnarray}
H_{\rm int} &=& 
g_{f}\int dx [\Psi^{\dagger}_{1\uparrow}\Psi_{1\uparrow}
\bar{\Psi}^{\dagger}_{1\downarrow}\bar{\Psi}_{1\downarrow}
+ \bar{\Psi}^{\dagger}_{2\uparrow}\bar{\Psi}_{2\uparrow}
\Psi^{\dagger}_{2\downarrow}\Psi_{2\downarrow}]
\nonumber\\
&+& g^{\prime}_{f}\int dx [\Psi^{\dagger}_{1\uparrow}\Psi_{1\uparrow}
\bar{\Psi}^{\dagger}_{2\uparrow}\bar{\Psi}_{2\uparrow}
+
\bar{\Psi}^{\dagger}_{1\downarrow}\bar{\Psi}_{1\downarrow}
\Psi^{\dagger}_{2\downarrow}\Psi_{2\downarrow} ]
\nonumber\\
&+& g_{u}
\int dx \{[\Psi^{\dagger}_{1\uparrow}(x)\Psi^{\dagger}_{1\uparrow}(x+a)
\bar{\Psi}_{1\downarrow}(x+a)\bar{\Psi}_{1\downarrow}(x) 
\nonumber\\
&+& 
\Psi^{\dagger}_{2\downarrow}(x)\Psi^{\dagger}_{2\downarrow}(x+a)
\bar{\Psi}_{2\uparrow}(x+a)\bar{\Psi}_{2\uparrow}(x) ]
\delta
+ h.c. \}
\nonumber\\
&+& g^{\prime}_{u}\int dx [\Psi^{\dagger}_{2\downarrow}
\Psi^{\dagger}_{1\uparrow}\bar{\Psi}_{1\downarrow}\bar{\Psi}_{2\uparrow}
\delta 
+ h.c.]
\nonumber\\
&+& g_{sf}\int dx [
\bar{\Psi}^{\dagger}_{2\uparrow}\Psi^{\dagger}_{1\uparrow}
\bar{\Psi}_{1\downarrow}\Psi_{2\downarrow}
+ \Psi^{\dagger}_{2\downarrow}\bar{\Psi}^{\dagger}_{1\downarrow}
\Psi_{1\uparrow}\bar{\Psi}_{2\uparrow}]
\label{eq:Hint}
\end{eqnarray}
by right movers (left-movers) $\Psi_{1\uparrow}$ and $\Psi_{2\downarrow}$ 
($\bar{\Psi}_{1\downarrow}$ and $\bar{\Psi}_{2\uparrow}$) with 
$\delta=\exp(-i4k_{F}x)$.  
$g_{f}$, $g^{\prime}_{f}$, 
$g_{u}$, $g^{\prime}_{u}$, and $g_{sf}$ 
denote the interaction constant of 
intra-edge forward, inter-edge forward, 
intra-edge umklapp, inter-edge umklapp,  
and inter-edge pseudospin-flip backward scattering, 
respectively. 
The corresponding Hamiltonian with parallel case can be derived 
by substituting $\bar{\Psi}_{2\downarrow}$ and $\Psi_{2\uparrow}$ for  
$\bar{\Psi}_{2\uparrow}$ and $\Psi_{2\downarrow}$. \par
Fermion operators are expressed by 
$\Psi_{1\uparrow}=\eta_{1\uparrow}\exp(-i\sqrt{4\pi}\phi_{1\uparrow})
/\sqrt{2\pi}$, 
$\bar{\Psi}_{1\downarrow}
=\bar{\eta}_{1\downarrow}
\exp(i\sqrt{4\pi}\bar{\phi}_{1\downarrow})/\sqrt{2\pi}$, 
$\bar{\Psi}_{2\uparrow}
=\bar{\eta}_{2\uparrow}\exp(i\sqrt{4\pi}\bar{\phi}_{2\uparrow})/\sqrt{2\pi}$, 
and $\Psi_{2\downarrow}=\eta_{2\downarrow}\exp(-i\sqrt{4\pi}\phi_{2\downarrow})/\sqrt{2\pi}$, 
with boson fields $\phi_{1\uparrow}$, $\bar{\phi}_{1\downarrow}$
$\bar{\phi}_{2\uparrow}$ and $\phi_{2\downarrow}$ \cite{Senechal}. 
$\eta_{1\uparrow}$, $\bar{\eta}_{1\downarrow}$, 
$\bar{\eta}_{2\uparrow}$ and ${\eta}_{2\downarrow}$ 
are Klein factors.  

Now we consider the bosonization of the Hamiltonians Eq.(\ref{eq:Hk}) and 
Eq.(\ref{eq:Hint})~\cite{Solyom}. 
By introducing 
$\varphi_{1}=\phi_{1\uparrow} + \bar{\phi}_{1\downarrow}$, 
$\varphi_{2}=\phi_{2\downarrow} + \bar{\phi}_{2\uparrow}$, 
$\theta_{1}=\phi_{1\uparrow} - \bar{\phi}_{1\downarrow}$, 
and
$\theta_{2}=\phi_{2\downarrow} - \bar{\phi}_{2\uparrow}$, 
the Hamiltonian $H$ can be given by 
\begin{eqnarray}
&&\frac{v}{2}
\int dx [(\frac{ \partial \varphi_{1}}{\partial x})^{2}
+ (\frac{ \partial \varphi_{2}}{\partial x})^{2} 
+
(\frac{ \partial \theta_{1}}{\partial x})^{2}
+ (\frac{ \partial \theta_{2}}{\partial x})^{2} 
]
\nonumber\\
&&
+ \frac{g_{f}}{4\pi}
\int dx [(\frac{ \partial \varphi_{1}}{\partial x})^{2}
+ (\frac{ \partial \varphi_{2}}{\partial x})^{2} 
-
(\frac{ \partial \theta_{1}}{\partial x})^{2}
- (\frac{ \partial \theta_{2}}{\partial x})^{2}]
\nonumber\\
&&
+ \frac{g^{\prime}_{f}}{2\pi}\int dx
[\frac{\partial \varphi_{1}}{\partial x}
\frac{\partial \varphi_{2}}{\partial x}
-
\frac{\partial \theta_{1}}{\partial x}
\frac{\partial \theta_{2}}{\partial x}]
\nonumber\\
&&
+\frac{g_{u}}{2\pi^{2}} \int
dx \{ \cos[4(\sqrt{\pi} \varphi_{1} -k_{F}x)]
+ \cos[4(\sqrt{\pi} \varphi_{2} -k_{F}x)]
\}
\nonumber\\
&&
+\frac{g^{\prime}_{u}}{2\pi^{2}}
\zeta
\int dx
\cos[\sqrt{4\pi}(\varphi_{1} + \varphi_{2})-4k_{F}x]
\nonumber\\
&&
-\frac{g_{sf}}{2\pi^{2}}
\zeta
\int dx
\cos[\sqrt{4\pi}(\varphi_{1}-\varphi_{2})]
\label{eq:bos1}
\end{eqnarray}
with $\zeta=\eta_{2\downarrow}\eta_{1\uparrow}
\bar{\eta}_{1\downarrow}\bar{\eta}_{2\uparrow}$.
In order to diagonalize the inter-edge forward 
scattering terms ($g^{\prime}_{f}$), we introduce the 
symmetric and antisymmetric combinations by 
$\varphi^{\prime}_{s} 
=(\varphi_{1} + \varphi_{2})/\sqrt{2}$, 
$\varphi^{\prime}_{a} 
=(\varphi_{1} - \varphi_{2})/\sqrt{2}$, 
$\theta^{\prime}_{s} 
=(\theta_{1} + \theta_{2})/\sqrt{2}$,
and 
$\theta^{\prime}_{a} 
=(\theta_{1} - \theta_{2})/\sqrt{2}$. 
Then, the above Hamiltonian Eq.(\ref{eq:bos1}) 
can be transformed into  
$H = H_0 + H_u + H_{u'} + H_{sf}$ with each of the 
term being given by 
\begin{eqnarray}
H_0&=&\int dx
\{
\frac{v_{1}}{2}[(\frac{\partial \varphi^{\prime}_{s}}{\partial x})^{2} +
K_{1}^{2} (\frac{\partial \theta^{\prime}_{s}}{\partial x})^{2}]
\nonumber \\
&+&
\frac{v_{2}}{2}[(\frac{\partial \varphi^{\prime}_{a}}{\partial x})^{2} +
K_{2}^{2} (\frac{\partial \theta^{\prime}_{a}}{\partial x})^{2}] \}
\label{eq:H0}, \\
H_u &=&
\frac{g_{u}}{\pi^{2}} \int
dx \cos(\sqrt{8\pi} \varphi^{\prime}_{s} -4k_{F}x)
\cos(\sqrt{8\pi}\varphi^{\prime}_{a}), 
\label{eq:Hu}
\\
H_{u'} &=&\frac{g^{\prime}_{u}}{2\pi^{2}}
\zeta
\int dx
\cos[\sqrt{8\pi}\varphi^{\prime}_{s}-4k_{F}x]
\label{eq:Hu'},
\\
H_{sf} &=& - \frac{g_{sf}}{2\pi^{2}}
\zeta
\int dx
\cos(\sqrt{8\pi}\varphi^{\prime}_{a})
\label{eq:Hsf}.
\end{eqnarray}
The dimensionless parameters 
$K_{1}$ and $K_{2}$ in Eq. (\ref{eq:H0}) are given by 
$K_{1}=\sqrt{\frac{2\pi v -(g_{f}+g^{\prime}_{f})}{2\pi v +(g_{f}+g^{\prime}_{f})}}$
and 
$K_{2}=\sqrt{\frac{2\pi v -(g_{f}-g^{\prime}_{f})}{2\pi v +(g_{f}-g^{\prime}_{f})}}$ with $v_{1}=v  + (g_{f}+g^{\prime}_{f})/(2\pi)$ and 
$v_{2}=v  + (g_{f} -g^{\prime}_{f})/(2\pi)$. 
Note that the difference between $K_{1}$ and $K_{2}$ is due to the 
presence of inter-edge forward scattering $g^{\prime}_{f}$. 
Now,  we consider the relevance/irrelevance 
of the nonlinear terms $H_{u}$, $H_{u'}$ and 
$H_{sf}$. 
These terms are relevant for  
$K_{1}+K_{2} < 1 $, $K_{1}<1 $ and $K_{2}<1$, respectively.  
For the single edge case, $K_{1}=K_{2}=K$, $H_{u}$ becomes relevant for $K<1/2$ \cite{wu2006}, 
which requires strong repulsive force.  
The inter-edge interactions $H_{u'}$ and $H_{sf}$ can be relevant more easily. 
The terms $H_{u}$ and $H_{u'}$ survive only in the commensurate case; 
$4k_{F}=G$ with a reciprocal lattice vector $G$. 
$H_{sf}$, on the other hand, exists even for the incommensurate case. \par

Let us first consider the incommensurate case, which is the most general case because 
the crossing energy of the helical edge modes are not necessarily at 
the center of the gap without the particle-hole symmetry, and also 
the Fermi energy is often determined by the details of the sample, $e.g.$, 
in-gap impurity levels and does not give the commensurate filling of the 
edge channels. 
In this case, only the inter-edge pseudospin-flip backward scattering $g_{sf}$ is relevant for  $K_{2}<1$, 
which is realized when the intra-edge repulsive forward scattering 
$g_{f}(>0)$ is stronger than the inter-edge repulsion $g^{\prime}_{f}(>0)$ for example. 
This relevant $g_{sf}$ term fixes $\varphi^{\prime}_{a}$ to open the gap for its fluctuation. 
The remnant massless modes are the conjugate pair 
$\varphi^{\prime}_{s}$ and $\theta^{\prime}_{s}$, which correspond to the 
correlated motion of the charge/spin between the two edges analogous to the 
bipolaron formation in spinfull ladder system \cite{Giamarchi,Nagaosa}. \par

We first focus on the generic phase diagram.  
It is determined by comparing the 
critical exponents of diverging correlation function of the 
following four order parameters, $i.e.$, 
(i)Intra-edge Density wave (DW), 
$\Psi_{1\uparrow}^{\dagger}\bar{\Psi}_{1\downarrow} 
\pm \Psi_{2\downarrow}^{\dagger}\bar{\Psi}_{2\uparrow}$; 
(ii)Inter-edge DW, 
$\Psi_{1\uparrow}^{\dagger}\bar{\Psi}_{2\uparrow} 
\pm \bar{\Psi}_{1\downarrow}^{\dagger}\Psi_{2\downarrow}$; 
(iii)
Intra-edge Superconductivity (SC)
$\Psi_{1\uparrow}\bar{\Psi}_{1\downarrow} 
\pm \Psi_{2\downarrow}\bar{\Psi}_{2\uparrow}$, 
and 
(iv)Inter-edge SC, 
$\Psi_{1\uparrow}\bar{\Psi}_{2\uparrow} 
\pm \bar{\Psi}_{1\downarrow}\Psi_{2\downarrow}$. 
Using the standard bosonization method, the correlation function of 
each of the above 
order parameters behaves as
\begin{equation} 
<O(x,\tau)O(0,0)> \sim [{\rm{max}} (\mid x \mid, v_{i=1,2} \mid \tau \mid)]^{-\nu}
\nonumber
\end{equation}
in the presence of gapless modes. 
Figure 2 shows the phase diagram in the plane of 
$g_{f}$ and $g^{\prime}_{f}$. The phase boundaries, which are
the straight lines $g_{f}= g^{\prime}_{f}$, $g_{f}= -g^{\prime}_{f}$
correspond to $K_{2}=1$, $K_{1}=1$, respectively. 
Namely, each region is characterized by 
(A) $K_{1}<1$, $K_{2}<1$,
(B) $K_{1}>1$, $K_{2}<1$,
(C) $K_{1}>1$, $K_{2}>1$,
and (D) $K_{1}<1$, $K_{2}>1$, respectively.
Table I and Table II explain the 
behaviors of the order parameter correlation functions 
for the incommensurate and commensurate cases, 
respectively.
For incommensurate case (Table I),
$H_{sf}$ is relevant and IECL is realized in regions A and B.
In this case, only the (i) Intra-edge DW or (iv) Inter-edge SC shows the 
critical power-law behavior, while the others are exponentially decaying
($"-"$ symbol in Table I). The most dominant order in each region is indicated
by the bold letters, i.e., (i)Intra-edge DW in region A, 
(ii)Inter-edge DW in region D, 
(iii)Intra-edge SC in region C, 
and (iv)Inter-edge SC in region B. \par
\begin{figure}[htb]
\begin{center}
\scalebox{0.8}{
\includegraphics[width=5cm,clip]{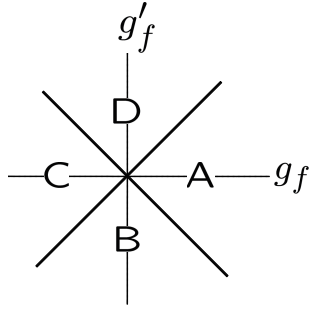}
}
\end{center}
\caption{
Phase diagram of two helical edge modes.
Intra-edge Density wave (DW), 
Inter-edge Superconductivity (SC), Intra-edge SC, and Inter-edge DW,
are most dominant in the region A, B, C, and D, respectively.  
}
\label{fig:2}
\end{figure}
\begin{center}
\begin{table}[h]
\begin{tabular}{|p{1.4cm}|p{1.1cm}|p{1.1cm}|p{2.4cm}|p{2.1cm}|}
\hline
Region & A & B & C & D \\ \hline
state & \multicolumn{2}{|c|}{IECL} & 
\multicolumn{2}{|c|}{TL} \\ \hline
Intra DW & ${\bm{K_{1}}}$ & $K_{1}$ & 
$K_{1}+K_{2}$ & $K_{1}+K_{2}$ \\ \hline
Inter DW & \multicolumn{2}{|c|}{$-$} &
$K_{1}+1/K_{2}$ & $\bm{K_{1}}+\bm{1/K_{2}}$ \\ \hline
Intra SC & \multicolumn{2}{|c|}{$-$} 
& $\bm{1/K_{1}}+\bm{1/K_{2}}$
& $1/K_{1} + 1/K_{2}$  \\ \hline
Inter SC
& $1/K_{1}$
& ${\bm{1/K_{1}}}$ 
& $1/K_{1} + K_{2}$
& $1/K_{1} + K_{2}$
\\ \hline
\end{tabular}%
\caption{
The relation between the most dominant order parameters and 
the exponents of correlation function for incommensurate case. 
$"-"$ denotes that the correlation function has an exponential dependence. 
The most dominant order in each region is indicated
by the bold letters. 
 }
\end{table}
\end{center}
We further study the physical properties of the IECL other than
the order parameters. 
First, we consider the stability of this state
against the inter-edge tunneling which has been 
neglected so far. The Hamiltonian describing this process is 
given by 
\begin{equation}
H_{t}=-\frac{2}{\pi}\int dx 
\{ t
\cos[\sqrt{2\pi}(\varphi^{\prime}_{s}+\theta^{\prime}_{a})] 
- t' 
\cos[\sqrt{2\pi}(\varphi^{\prime}_{a}+\theta^{\prime}_{a})] 
\}
\end{equation}
where $t$ and $t'$ are transfer integrals 
without and with pseudospin-flip scattering, respectively. 
It is evident that this perturbation is 
irrelevant since $\varphi^{\prime}_{a}$ is fixed in IECL. \par
Next, we consider the magnetic properties of IECL. 
In this state, the spin-spin correlation function 
has two contributions, 
$<S(x=0,\tau)S(x=0,0)> \sim \frac{1}{\tau^{2}} + (\frac{1}{\tau})^{K_{1}}$. 
The Fourier transform with respect to the imaginary time $\tau$ thus 
leads to the spin susceptibility 
$\chi_{s} \sim T + T^{K_{1}-1}$ as a function of temperature 
$T$~\cite{Giamarchi}. 
Then the resulting temperature dependence of NMR relaxation rate $T_{1}$
is given by 
$\frac{1}{T_{1}} \sim T + T^{K_{1}-1}$. 
It should be remarked that $1/T_{1}$ has a divergent $T$
dependence as $T \to 0$ for $K_{1}<1$, where Intra-edge DW state 
is dominant realized in region A. 
This anomalous temperature dependence of $T_{1}$ \cite{NMR} 
can be detected by experiments of NMR. \par
The Josephson coupling through IECL shows also a novel property. 
Here, we consider the IECL with length $d$ is sandwiched by 
two superconductors S$_{1}$ and S$_{2}$ \cite{Josephson}. 
The Josephson current $I_{J}$ is given by 
$I_{J}=-2e \partial F(\chi) / \partial \chi$ with phase difference $\chi$. 
The phase dependent part of the free energy is proportional to 
$F(\chi) \propto
{\rm Re}[t_{1}^{2}(t_{2}^{*})^{2}e^{-i\chi}
\int^{\beta}_{0} d\tau \Pi(d,\tau)]$
with $\Pi(d,\tau)=<O(0,0)O(d,\tau)>$,
$O(0,0)= \Psi_{1\uparrow}\bar{\Psi}_{2\uparrow} 
\pm \bar{\Psi}_{1\downarrow}\Psi_{2\downarrow}$ 
and  $\beta=1/T$. 
$t_{1}(t_{2})$ is the transmission amplitude at 
IECL/S$_{1}$(S$_{2}$) interface. 
Following the discussion by Fazio $et$ $al.$ \cite{Josephson}, 
the resulting $I_{J}$ at zero temperature 
is proportional to $d^{-1/K_{1} + 1}$ for $K_{1}<1$. 
This is quite unusual that Josephson current is hugely enhanced as compared to 
conventional spinfull Luttinger chain where $I_{J}$ is proportional to 
$d^{-1/K}$. 
\begin{center}
\begin{table}[h]
\begin{tabular}{|p{1.5cm}|p{1.2cm}|p{1.2cm}|p{2.4cm}|p{1.6cm}|}
\hline
Region & A & B & C & D \\ \hline
state & DW & IECL & TL & IECL \\ \hline
Intra DW & {\bf Order} & $K_{1}$ & 
$K_{1}+K_{2}$ & $K_{2}$ \\ \hline
Inter DW & $-$ & $-$ & 
$K_{1}+1/K_{2}$ & ${\bm{1/K_{2}}}$ \\ \hline
Intra SC & $-$ & $-$ & 
$\bm{1/K_{1}} + \bm{1/K_{2}}$
& $-$  \\ \hline
Inter SC
& $-$
& ${\bm{1/K_{1}}}$ 
& $1/K_{1} + K_{2}$
& $-$
\\ \hline
\end{tabular}%
\caption{
Similar Table to Table I for the commensurate case. 
}
\end{table}
\end{center}
Now we turn to the commensurate case (Table II). 
The term $H_{u'}$ [Eq. (\ref{eq:Hu'})] fixes the phase $\varphi'_{s}$ for 
$K_{1}>1$ which is realized in regions A and D. 
The Intra DW phase is stabilized in the region A and order parameter 
has a nonzero value, i.e., long-range ordered state at zero temperature. 
The Inter DW is also stabilized and the resulting $\nu$ 
becomes $1/K_{2}$ in the region D. 
On the other hand, 
the exponents $\nu$'s in the regions B and C do not change from the incommensurate case, since the umklapp terms is irrelevant there. 
Note that the most dominant order parameter in each 
region does not change from the incommensurate case.
\par
In the parallel helical edge modes (Fig.1(b)),
we can obtain essentially the same results as the anti-parallel
case discussed thus far. 
The only difference is that the corresponding operators 
for (ii)Inter-edge charge density wave (CDW)/ spin density wave (SDW) 
and (iv)Inter-edge SC are 
given as 
$\Psi_{1\uparrow}^{\dagger}\bar{\Psi}_{2\downarrow} 
\pm \bar{\Psi}_{1\downarrow}^{\dagger}\Psi_{2\uparrow}$ 
and 
$\Psi_{1\uparrow}\bar{\Psi}_{2\downarrow} \pm \bar{\Psi}_{1\downarrow}
\Psi_{2\uparrow}$. 
This means that the opposite pseudospin 
electron operators appear in the parallel case 
instead of the same pseudospin ones in the anti-parallel 
case. \par

Finally, we compare the present IECL with 
the one-dimensional liquids already known.
The single edge mode case is reproduced by setting $g^{\prime}_{f}=0$, 
$g^{\prime}_{u}=0$ and $g_{sf}=0$. Then  $K_{1}=K_{2}=K$ is satisfied. 
The intra-edge umklapp scattering $g_{u}$ becomes relevant 
for $K<1/2$ \cite{wu2006}. 
The exponent of correlation function $\nu$ of 
Intra-edge SC and Intra-edge DW 
are given by $2/K$ and $2K$, which shows weaker correlation compared 
with those in Table I and II.
For spinfull one-dimensional TL liquid, 
the $\nu$ of Intra-edge SC and Intra-edge CDW/SDW 
are given by $1/K_{\rho} + K_{\sigma}(1/K_{\sigma})$ and 
$K_{\rho} + K_{\sigma}(1/K_{\sigma})$~\cite{Giamarchi}.   
As compared to this spinfull one-dimensional TL liquid, 
the density wave (SDW/CDW) and 
superconducting correlations both of which 
are more strongly diverging. 
The spin gap phase is 
realized in Luther-Emery liquid~\cite{Giamarchi,Rashba}. 
It is known that the $\nu$ of the intra-edge SC and the intra-edge CDW 
are given by $1/K$ and $K$, respectively.  
Although these exponents are similar to those in IECL,  
IECL  is very much different 
since there is no spin gap.
This difference comes from the fact that 
the combination 
$\phi_{1 \uparrow}+\bar{\phi}_{2 \uparrow}
- \phi_{2 \downarrow}- \bar{\phi}_{1 \downarrow}$
is fixed in Luther-Emery liquid while a different one
$\phi_{1 \uparrow}-\bar{\phi}_{2 \uparrow}
- \phi_{2 \downarrow}+ \bar{\phi}_{1 \downarrow}$
is fixed in IECL. 
The present IECL is also different from spinfull two-leg ladder, 
where the $\nu$ of the inter-ladder $d$-wave pairing 
can become $1/(2K)$ \cite{Spingap}. 
However, there opens the spin gap also in this state,
in sharp contrast to the present IECL.

 In this letter, we have proposed an Inter-edge correlated 
liquid (IECL)  for the two helical edge modes of 
quantum spin Hall systems.  
IECL  is realized most probably in the incommensurate case, 
showing the density wave and 
superconducting correlations both of which 
are more strongly diverging compared with the one-dimensional spinfull 
interacting electrons. 
Furthermore, there is no spin gap in this liquid state, in sharp contrast to 
the Luther-Emery liquid or the spin gap state in two-leg ladder system. 
This state can be clearly identified experimentally by NMR and
Josephson junctions. \par

This work is partly supported by the
Grant-in-Aids from under the Grant No.\ 20654030, 
No. 19048015 and No. 19048008 from 
MEXT, Japan, and NTT basic research laboratories.


\end{document}